\documentclass[%
 aip,
 jcp,%
 amsmath,amssymb,
preprint,%
 floatfix,%
]{revtex4-2}
\usepackage[T1]{fontenc}
\usepackage{microtype}
\usepackage{graphicx}
\usepackage{bm}
\usepackage{booktabs}
\usepackage[version=4]{mhchem}
\usepackage{siunitx}
\DeclareSIUnit\angstrom{\text{\AA}}
\usepackage[colorlinks=true,linkcolor=blue,citecolor=blue,urlcolor=blue]{hyperref}
\usepackage[nameinlink]{cleveref}

\crefname{table}{Table}{Tables}
\Crefname{table}{Table}{Tables}
\crefname{figure}{Fig.}{Figs.}
\Crefname{figure}{Fig.}{Figs.}
\crefname{equation}{Eq.}{Eqs.}
\Crefname{equation}{Eq.}{Eqs.}

\makeatletter
\def\frontmatter@thefootnote{*}
\def\frontmatter@makefnmark{\@textsuperscript{\normalfont\@thefnmark\if*\@thefnmark\else)\fi}}
\makeatother

\begin{document}
\title{Nuclear quantum effects enhance diffusion in supercooled ammonia through accelerated pyramidal inversion}

\author{Minwoo Kim}
\author{Hido Woo}
\author{Taeyong Park}
\affiliation{Department of Chemical and Biological Engineering, Institute of Chemical Processes, Seoul National University, Seoul 08826, Republic of Korea}
\author{Ji Woong Yu}
\email[Corresponding Authors: ]{jiwoongs1492@ajou.ac.kr (J.W.Y.); wblee@snu.ac.kr (W.B.L.)}
\affiliation{School of Frontier Sciences, Ajou University, Suwon 16499, Republic of Korea}
\affiliation{Department of Chemistry, Ajou University, Suwon 16499, Republic of Korea}
\author{Won Bo Lee}
\email[Corresponding Authors: ]{jiwoongs1492@ajou.ac.kr (J.W.Y.); wblee@snu.ac.kr (W.B.L.)}
\affiliation{Department of Chemical and Biological Engineering, Institute of Chemical Processes, Seoul National University, Seoul 08826, Republic of Korea}
\affiliation{School of Transdisciplinary Innovations, Seoul National University, Seoul 08826, Republic of Korea}

\date{\today}

\begin{abstract}
In liquid ammonia, an imbalance between available hydrogen-bond donor and acceptor sites limits water-like tetrahedral connectivity, leaving sparse and short-lived associations within a densely packed liquid. We ask whether quantum-sensitive local rearrangements can nevertheless contribute appreciably to diffusion in this weakly connected liquid. We compare classical and thermostatted ring polymer molecular dynamics from 170 to 250~\unit{K} using an r$^2$SCAN-trained machine learning force field. Ammonia can change its molecular geometry as the nitrogen atom passes through the plane of the three hydrogen atoms, a motion known as pyramidal inversion. Nuclear quantum effects increase the inversion rate and reduce its apparent activation energy. Inversion events are accompanied by transient reductions in hydrogen-bond coordination and local density within the first solvation shell. Although nuclear quantum effects lower density and viscosity throughout the studied range, diffusion remains only weakly affected in the warmer liquid. The quantum enhancement of diffusion emerges below approximately 210~\unit{K} and reaches 18\% at 170~\unit{K}, tracking the enhancement of cage escape. These observations support an inversion-assisted contribution to diffusion, in which a spatially localized, quantum-sensitive internal motion couples to cage relaxation without a persistent tetrahedral network.
\end{abstract}

\maketitle

\section{Introduction}
Simple liquids and network-forming complex liquids represent contrasting limits of structural organization. In simple liquids, packing and predominantly isotropic interactions provide a useful reference for structure and dynamics\cite{hansen2013theory,Ingebrigtsen2012what}. In water and liquid silicon, directional bonding instead favors open tetrahedral configurations that compete with dense packing\cite{Tanaka2002Simple,russo2014understanding}. Competition between these local environments provides a structural basis for thermodynamic and transport anomalies and their dependence on tetrahedrality\cite{Morishita2005Anomalous,russo2018waterlike}. Weakly associating liquids occupy an intermediate regime, in which specific directional attractions supplement ordinary packing and dispersion\cite{Wertheim1984fluidsI,Wertheim1986fluidsIII}. They depart from the simple-liquid reference while remaining distinct from the strongly network-forming limit.

Liquid ammonia exemplifies this intermediate regime through its sparse and transient hydrogen bonds (HBs). Each molecule has three potential HB donor groups but a single lone-pair acceptor region, in contrast to the balanced nominal donor and acceptor valences of water. Neutron diffraction reveals a less pronounced HB signature than in water\cite{ricci1995microscopic}, while first-principles simulations report approximately one accepted HB per molecule with a lifetime near 0.1~\unit{ps}\cite{krishnamoorthy2022hydrogen}. This combination of dense packing and rapidly exchanging HBs motivates examining how changes in molecular geometry are coupled to rearrangement of the surrounding liquid.

Pyramidal inversion is one such change in geometry, in which the nitrogen atom passes through the plane of the three hydrogen atoms, interconverting two pyramidal configurations. For an isolated molecule, these configurations have the same energy\cite{spirko1983vibrational,rajamaki2003vibrational}. In the liquid, consider holding the neighboring molecules fixed while the central molecule inverts. The resulting changes in intermolecular contacts can make the inverted configuration less favorable, even though the neighbors have not moved. Rearrangement of the neighbors can accommodate the altered geometry, suggesting a coupling between inversion and solvation-shell relaxation. 

Path integral molecular dynamics has been used to examine nuclear quantum effects (NQEs) in liquid ammonia\cite{diraison1999simulation}. Neutron-scattering measurements interpreted with quantum simulations have further established the relevance of NQEs to anharmonic molecular motion in condensed ammonia\cite{linker2024neutron}. We ask whether quantum effects on pyramidal inversion are also reflected in translational diffusion through rearrangement of the surrounding liquid.

Any such contribution should depend on the relaxation timescale of the surrounding liquid. When thermally driven rearrangements are rapid, a short-lived, localized perturbation of the solvation shell may contribute little to long-time diffusion. On cooling, however, translational dynamics becomes increasingly suppressed by caging, a phenomenon that also occurs in simple glass formers without a persistent bonding network\cite{tong2018revealing,tong2019structural}. In this regime, local HB loss accompanying inversion can loosen the cage and facilitate escape. A higher inversion rate with quantum nuclei would increase the frequency of these local relaxation events relative to the classical liquid. Their contribution to diffusion can consequently become appreciable as thermal relaxation is suppressed, even if the average local structure changes only modestly.

Testing this possibility requires both an accurate description of the anharmonic inversion pathway and its intermolecular environment, and a dynamical treatment that incorporates NQEs. Four of the five conventional force fields tested here show no inversion events during the sampled trajectories, although some reproduce selected bulk properties accurately. We use the r$^2$SCAN-trained machine learning force field (MLFF) developed in our previous work\cite{kim2026bottom}, which describes liquid structure, density, viscosity, and diffusion with accuracy comparable to the best conventional models tested while retaining spontaneous inversion. For classical nuclei, direct event counting and enhanced sampling yield consistent apparent inversion activation energies. The computational efficiency of the MLFF makes nanosecond sampling of rare inversions feasible for hundreds of molecules, including the multiple replicas needed for ring polymer simulations.

To examine how NQEs modify inversion and transport, we compare classical molecular dynamics (MD) with thermostatted ring polymer molecular dynamics (TRPMD) on the same MLFF potential energy surface from 170 to 250~\unit{K}. In the underlying ring polymer molecular dynamics (RPMD) framework, each nucleus is represented by a ring polymer of harmonically connected beads, incorporating quantum nuclear statistics while providing an approximate description of real-time dynamics\cite{craig2004quantum}. In TRPMD, only the internal ring polymer modes are thermostatted, leaving the centroid unthermostatted\cite{rossi2014how}. We characterize inversion energetics and kinetics and then compare distributions of HB coordination, local density, and structural order with the changes surrounding individual inversion events. This comparison separates the liquid's average structural organization from the transient rearrangements coupled to inversion. The examined distributions remain unimodal, yet individual events are accompanied by HB loss and reduced local density within the first solvation shell. NQEs increase the inversion rate relative to classical dynamics, while their small contribution to diffusion in the warmer liquid becomes positive as caging develops on cooling. The corresponding enhancement of cage escape tracks that of diffusion, supporting inversion as a local structure breaker that contributes to translational relaxation in supercooled ammonia.

\section{Methods}
\subsection{Force fields}
Eckl\cite{eckl2008optimised} and TraPPE\cite{zhang2010development} are rigid non-polarizable models and were propagated as rigid bodies with a time-reversible symplectic integrator\cite{kamberaj2005time}. The parameters of Eckl were fitted to the saturated liquid density, vapor pressure, and enthalpy of vaporization, and those of TraPPE to the vapor--liquid coexistence curve. OPLS-AA\cite{jorgensen1996development,rizzo1999opls} and OpenFF\cite{qiu2021development,boothroyd2023development} Sage version 2.3.0\cite{wang2026developing} are flexible non-polarizable models with harmonic bond and angle terms. 

The non-bonded parameters of OPLS-AA were fitted to the densities and heats of vaporization of organic liquids, and its amine parameters were refined against pure-liquid properties and gas-phase HB strengths and subsequently validated. OpenFF assigns parameters by chemical perception. Its valence terms were fitted to quantum-chemical data, and its non-bonded terms to the densities and enthalpies of mixing of organic liquids. AMOEBA\cite{ren2003polarizable,ren2011polarizable} was used with the amoeba09 parameter set. Its atomic multipoles were derived from first-principles distributed multipole analysis, and its non-bonded parameters were fitted to gas-phase dimer energies and to the densities and heats of vaporization of liquids. 

The deep potential molecular dynamics (DPMD) model of ammonia was taken from our previous work on first-principles molecular thermodynamics\cite{kim2026bottom}, which describes its training data, network architecture, and validation against first-principles calculations. The model was trained on data from the r$^2$SCAN exchange-correlation functional. The same model propagated with TRPMD to include NQEs is denoted TRP-DPMD.

\subsection{Molecular dynamics simulations}
All systems contained 512 \ce{NH3} molecules in a cubic box with periodic boundary conditions, and the time step was 0.5~\unit{fs}. In the classical simulations, temperature was controlled with a Nos\'e--Hoover thermostat with a time constant of 0.5~\unit{ps}. Pressure in the isothermal--isobaric (NPT) ensemble was controlled with a Nos\'e--Hoover barostat with a time constant of 5~\unit{ps}.\cite{nose1984unified,hoover1985canonical}

For the force field validation, the radial distribution functions (RDFs), viscosity, and diffusion coefficient were obtained from 10~\unit{ns} simulations in the canonical (NVT) ensemble at 213~\unit{K} and 0.71365~\unit{g/cm^3}, the density at this temperature from the NIST Chemistry WebBook\cite{nist}. The density was obtained from NPT simulations at 213~\unit{K} and 1.0~\unit{bar}. For TRP-DPMD, only the RDFs were computed at 213~\unit{K}, and its density, viscosity, and diffusion coefficient at this temperature were interpolated between the 210 and 215~\unit{K} values of the temperature series described below.

For the temperature dependence, densities were obtained in the NPT ensemble at 1.0~\unit{bar} from 170 to 250~\unit{K} after 1~\unit{ns} of equilibration, and all other quantities were obtained in the NVT ensemble at the corresponding density. Production runs lasted 2.0~\unit{ns} and, because the inversion becomes rare at low temperatures, were extended with decreasing temperature up to 6.0~\unit{ns} at 170~\unit{K}. Each quantity was averaged over seven independent DPMD runs and two independent TRP-DPMD runs started from different initial configurations, and the reported uncertainties are standard errors over the runs.

Eckl, TraPPE, OPLS-AA, OpenFF, and DPMD were run in LAMMPS\cite{thompson2022lammps}. For the conventional force fields, Lennard-Jones interactions were cut off at 12~\unit{\angstrom} and electrostatic interactions were evaluated with the particle--particle particle--mesh method at a relative accuracy of $10^{-5}$. AMOEBA was run in OpenMM\cite{eastman2017openmm} with the parameter file in Tinker format, a van der Waals cutoff of 12~\unit{\angstrom}, and particle mesh Ewald electrostatics\cite{darden1993particle,essmann1995smooth} with a real-space cutoff of 8~\unit{\angstrom}.

\subsection{Thermostatted ring polymer molecular dynamics}
NQEs were included with RPMD\cite{craig2004quantum} in its thermostatted form\cite{rossi2014how}, using the \texttt{fix pimd/langevin} command of LAMMPS\cite{li2026fix} with 32 beads per atom. The equations of motion were integrated in the normal-mode representation with the OBABO splitting and a time step of 0.5~\unit{fs}. The internal modes were thermostatted with the local path integral Langevin equation thermostat\cite{ceriotti2010efficient} using the optimal damping times of that work. The centroid mode was left effectively unthermostatted by setting its damping time orders of magnitude longer than the simulation time, so that the centroid follows the ring polymer dynamics without friction\cite{rossi2014how}. RDFs of TRP-DPMD were computed within each bead replica and averaged over the 32 replicas. Centroid coordinates were used for the trajectory-based inversion and displacement analyses.

\subsection{Reference data}
The simulated structure was compared with the RDFs of liquid ammonia at 213~\unit{K} from neutron diffraction with isotopic substitution\cite{ricci1995microscopic}. Reference densities and viscosities were taken from the NIST Chemistry WebBook\cite{nist} for the liquid between the melting and boiling points at 1.0~\unit{bar}. The reference diffusion coefficient was obtained from the Arrhenius fit of the NMR data of O'Reilly \emph{et al.}\cite{oreilly1973self}, $D = 4.5\times10^{-3}\exp(-0.0898~\unit{eV}/k_{\mathrm{B}}T)$~\unit{cm^2/s}, with $k_{\mathrm{B}}$ the Boltzmann constant and $T$ the temperature.

\subsection{Structural analysis}
RDFs were computed from the MD trajectories, and their peak positions and heights were obtained by quadratic interpolation around each peak maximum. The search ranges for the intermolecular peaks are given in Table~S2 of the supplementary material. The incremental probability density function, $g^{(i)}_{\mathrm{NH}}(r)$, is the distribution of the distance from a nitrogen atom to its $i$-th nearest hydrogen atom\cite{sharma2018born,sharma2018nature,zhuang2022hydration,avula2023understanding,kim2024anomalous}. Two molecules were regarded as hydrogen bonded when an intermolecular \ce{N\bond{...}H} distance between them was shorter than 2.4~\unit{\angstrom}\cite{krishnamoorthy2022hydrogen}.

The local density, $\rho_{\mathrm{loc}}$, of a molecule is the inverse of the volume of the Voronoi cell of its nitrogen atom. The local structural index\cite{russo2014understanding}, $\zeta$, was adapted to ammonia as the difference between the \ce{N\bond{-}N} distance to the nearest molecule that is not hydrogen bonded to the central molecule and the \ce{N\bond{-}N} distance to the farthest molecule that is hydrogen bonded to it. Molecules with no hydrogen-bonded neighbor were excluded from the evaluation of $\zeta$.

\subsection{Transport properties}
The viscosity was obtained from the Green--Kubo relation\cite{green1954markoff,kubo1957statistical} by integrating the autocorrelation function of the off-diagonal pressure tensor components over blocks of 5~\unit{ps} with a maximum lag of 3~\unit{ps}. The integral was estimated with the stable autocorrelation integral estimator (STACIE)\cite{toraman2025stable}, which fits the low-frequency part of the power spectrum up to a cutoff frequency selected by an information criterion.

The diffusion coefficient in the periodic box, $D_L$, was obtained from the slope of the mean square displacement of the nitrogen atoms in its linear regime and corrected for finite-size effects with the Yeh--Hummer relation\cite{yeh2004system} using the viscosity, $\eta_0$, and box length, $L$, of the same simulation, as
\begin{equation}
    D = D_L + 2.837297\,k_{\mathrm{B}}T/(6\pi\eta_0 L).
    \label{eq:yhrelation}
\end{equation}

\subsection{Inversion analysis}
The inversion coordinate, $h$, of a molecule is the signed distance of the nitrogen atom from the plane of its three hydrogen atoms. The hydrogen labels are kept fixed, so $h$ is odd under inversion, and the pyramidal minima lie at $h \approx \pm0.40$~\unit{\angstrom}. With $\mathbf{r}_{\mathrm{N}}$ and $\mathbf{r}_{\mathrm{H}_i}$ denoting the positions of the nitrogen atom and of the three hydrogen atoms of the molecule, the unit normal of the plane, $\hat{\mathbf{n}}$, the centroid of the hydrogen atoms, $\bar{\mathbf{r}}_{\mathrm{H}}$, and $h$ are given by
\begin{equation}
    \hat{\mathbf{n}} = \frac{(\mathbf{r}_{\mathrm{H}_2} - \mathbf{r}_{\mathrm{H}_1}) \times (\mathbf{r}_{\mathrm{H}_3} - \mathbf{r}_{\mathrm{H}_1})}{\lVert (\mathbf{r}_{\mathrm{H}_2} - \mathbf{r}_{\mathrm{H}_1}) \times (\mathbf{r}_{\mathrm{H}_3} - \mathbf{r}_{\mathrm{H}_1}) \rVert}, \label{eq:normal}
\end{equation}
\begin{equation}
    \bar{\mathbf{r}}_{\mathrm{H}} = \frac{1}{3} \sum_{i=1}^{3} \mathbf{r}_{\mathrm{H}_i}, \label{eq:centroid}
\end{equation}
\begin{equation}
    h = \hat{\mathbf{n}} \cdot \left( \mathbf{r}_{\mathrm{N}} - \bar{\mathbf{r}}_{\mathrm{H}} \right). \label{eq:h}
\end{equation}

Inversion events in unbiased simulations were detected with the hop function, $H$\cite{candelier2009building,park2026beyond}, evaluated on $h$, as defined in the Results, with a threshold of 0.15~\unit{\angstrom^2}. The rate constant, $k$, is the number of events per molecule per unit time, and the activation energy was obtained from an Arrhenius fit of $k$.

Free energy profiles along the inversion coordinate were obtained with on-the-fly probability enhanced sampling (OPES)\cite{invernizzi2020rethinking} as implemented in PLUMED\cite{tribello2014plumed,plumed2019promoting}. The bias was applied to $h$ of a single tagged molecule in the liquid, with a barrier estimate of 0.31~\unit{eV}, a kernel width of 0.05~\unit{\angstrom} bounded below by 0.005~\unit{\angstrom}, a deposition stride of 200 steps, and a kernel compression threshold of 1.0. Profiles were computed from 180 to 240~\unit{K} in steps of 10~\unit{K} at the corresponding densities, and each profile is the average over three independent runs. 

The gas-phase profile was obtained for an isolated molecule with the same settings. Rate constants were obtained from the profiles by transition state theory.

Minimum energy paths of the inversion were obtained with the nudged elastic band (NEB) method as implemented in ASE\cite{larsen2017atomic}, using the same MLFF, for an isolated molecule and for a molecule in the liquid with all other molecules held fixed at their positions in the initial configuration. The initial image was the pyramidal minimum, relaxed with the FIRE optimizer\cite{bitzek2006structural} to a maximum force of 0.02~\unit{eV/\angstrom}, and the final image was obtained by reflecting the nitrogen atom through the plane of its three hydrogen atoms and relaxing again with the same criterion. The path consisted of 30 intermediate images generated by linear interpolation between the two endpoints and connected by springs of 0.1~\unit{eV/\angstrom^2}. It was first converged to a maximum force of 0.10~\unit{eV/\angstrom} without climbing and then refined with the climbing-image method\cite{henkelman2000climbing} to 0.03~\unit{eV/\angstrom}.

\subsection{Analysis of the inversion impact}
The number of HBs and the local density of the inverting molecule were averaged over all events as functions of time relative to the inversion within $\pm$20~\unit{ps}. The same quantities were resolved by the distance from the inverting molecule between 3 and 11~\unit{\angstrom} and expressed relative to the bulk averages.

Mean square displacements, $\langle\Delta r^{2}\rangle$, were computed separately for molecules that inverted within the time window and for those that did not. The cage was characterized by the logarithmic slope of the mean square displacement, $\beta(t) = \mathrm{d}\ln\langle\Delta r^{2}\rangle/\mathrm{d}\ln t$. The time at which $\beta$ is minimal between 0.5 and 20~\unit{ps}, $t^{*}$, marks the most subdiffusive regime, and the cage length, $\ell$, is the root mean square displacement of the nitrogen atoms over $t^{*}$. The observation time, $t^{*}$, and cage length, $\ell$, were determined from the TRP-DPMD run and applied identically to DPMD and TRP-DPMD at each temperature. At temperatures for which the minimum of $\beta$ exceeded 0.95, no cage was resolved by this criterion. For those temperatures, the values of $t^{*}$ and $\ell$ of the highest temperature at which a cage was resolved were retained. A molecule was regarded as having escaped its cage when its displacement satisfied the criterion, $|\Delta r(t^{*})| > 1.5\,\ell$.

\section{Results and Discussion}
\subsection{Force field validation}
We benchmarked the MLFF against five conventional force fields spanning rigid non-polarizable, flexible non-polarizable, and flexible polarizable descriptions of liquid ammonia. Eckl\cite{eckl2008optimised} and TraPPE\cite{zhang2010development} are rigid non-polarizable, OPLS-AA\cite{jorgensen1996development,rizzo1999opls} and OpenFF\cite{qiu2021development,boothroyd2023development,wang2026developing} are flexible non-polarizable, and AMOEBA\cite{ren2003polarizable,ren2011polarizable} is flexible and polarizable (see Methods for their parametrization). We compared the RDFs for structure, the density for thermodynamics, and the viscosity and diffusion coefficient for transport with reference data from NIST and experiment. At 213~\unit{K}, the RDFs and the transport properties with classical nuclei were evaluated in the NVT ensemble at the NIST density, while direct density calculations used the NPT ensemble at 1~\unit{bar}. The TRP-DPMD RDFs were evaluated at the same reference density. The density and transport properties of TRP-DPMD at 213~\unit{K} were instead interpolated from the temperature series, for which transport was evaluated at the corresponding equilibrium densities.

\subsubsection{Liquid structure}
\begin{figure}[tbp]
    \centering
    \includegraphics[width=3.37in]{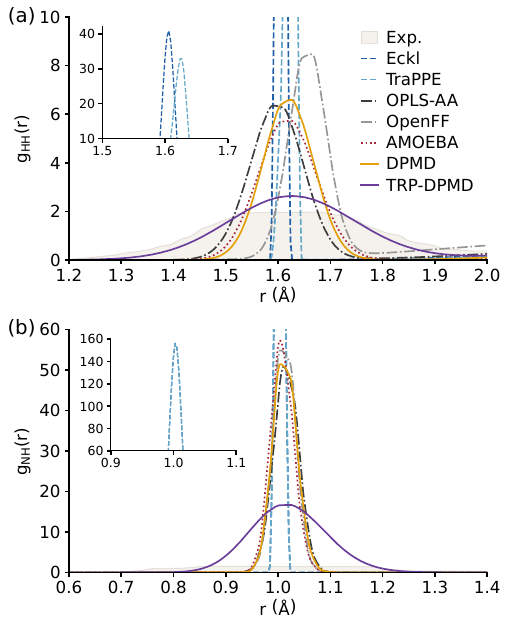}
    \caption{Intramolecular radial distribution functions, $g(r)$, of liquid ammonia. (a) Hydrogen--hydrogen, $g_{\mathrm{HH}}(r)$, and (b) nitrogen--hydrogen, $g_{\mathrm{NH}}(r)$. Insets magnify the first peak. Shaded areas (Exp.) are the neutron diffraction data of Ricci \emph{et al.}\cite{ricci1995microscopic}. All simulations were performed in the canonical ensemble at 213~\unit{K} and 0.71365~\unit{g/cm^3}. Eckl\cite{eckl2008optimised} and TraPPE\cite{zhang2010development} are rigid non-polarizable models, OPLS-AA\cite{jorgensen1996development,rizzo1999opls} and OpenFF\cite{qiu2021development,boothroyd2023development} are flexible non-polarizable models, and AMOEBA\cite{ren2003polarizable,ren2011polarizable} is a flexible polarizable model. DPMD and TRP-DPMD denote classical MD and TRPMD simulations, respectively, using the same MLFF developed in our previous work\cite{kim2026bottom}.}
    \label{fig:rdf_intra}
\end{figure}

We first assessed the liquid structure through the RDFs, $g_{ij}(r)$, which give the density of atoms $j$ at a distance, $r$, from an atom $i$ relative to the bulk density. Peak positions and heights were compared with the neutron diffraction data of Ricci \emph{et al.}\cite{ricci1995microscopic}, treating the intramolecular and intermolecular regions separately. Every model reproduced the positions of the two intramolecular peaks, the \ce{N\bond{-}H} bond in $g_{\mathrm{NH}}(r)$ and the \ce{H\bond{-}H} separation within a molecule in $g_{\mathrm{HH}}(r)$ (\cref{fig:rdf_intra} and Table~S1). The peak heights, by contrast, spanned an order of magnitude. Molecular stoichiometry fixes the coordination integral $4\pi n_j\int r^2 g_{ij}^{\mathrm{intra}}(r)\,\mathrm{d}r$, where $n_j$ is the bulk number density of species $j$. At comparable peak positions, the lower peak heights reflect broader distributions of intramolecular distances. The rigid models Eckl and TraPPE gave the sharpest peaks, because their geometry fixes both the bond lengths and the bond angle. The flexible models with classical nuclei, OPLS-AA, OpenFF, AMOEBA, and DPMD, gave heights within a factor of two of one another, but all remained far sharper than the experimental peaks.

We included NQEs through TRPMD, in which a ring polymer of 32 copies of the system connected by harmonic springs reproduces the equilibrium properties of the quantum nuclei (see Methods). TRP-DPMD broadened the bead-averaged intramolecular distance distributions and shifted the \ce{N\bond{-}H} peak slightly outward (\cref{fig:rdf_intra}). These changes reflect the effect of nuclear quantum fluctuations on the intramolecular distance distributions.

\begin{figure}[tbp]
    \centering
    \includegraphics[width=3.37in]{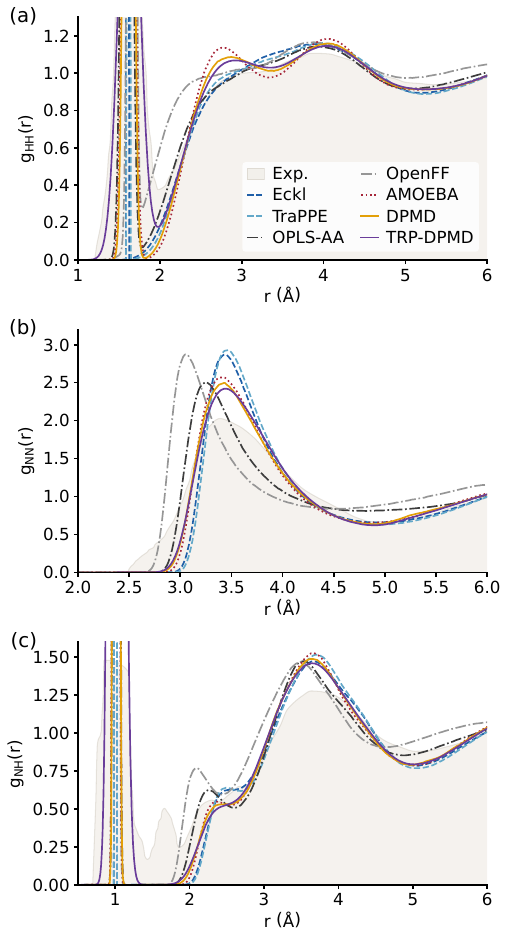}
    \caption{Radial distribution functions, $g(r)$, of liquid ammonia. (a) Hydrogen--hydrogen, $g_{\mathrm{HH}}(r)$, (b) nitrogen--nitrogen, $g_{\mathrm{NN}}(r)$, and (c) nitrogen--hydrogen, $g_{\mathrm{NH}}(r)$. Shaded areas (Exp.) are the neutron diffraction data of Ricci \emph{et al.}\cite{ricci1995microscopic}. Simulation conditions and model labels are the same as in \cref{fig:rdf_intra}.}
    \label{fig:rdf_inter}
\end{figure}

Unlike the intramolecular structure, the intermolecular structure diverged among the models below 5~\unit{\angstrom} in all three partial RDFs, and the deviations sorted the models into three groups (\cref{fig:rdf_inter} and Table~S2). In peak position, the flexible non-polarizable models OPLS-AA and OpenFF shifted the first peak of $g_{\mathrm{NN}}(r)$ and the main peak of $g_{\mathrm{NH}}(r)$ between 3 and 4~\unit{\angstrom} to shorter distances (\cref{fig:rdf_inter}b and c). Every other model placed these peaks close to the experimental positions. In $g_{\mathrm{HH}}(r)$ (\cref{fig:rdf_inter}a), AMOEBA, DPMD, and TRP-DPMD gave an inner and an outer peak, whereas Eckl, TraPPE, OPLS-AA, and OpenFF gave only the outer one, as did the experiment.

In peak height, every model overestimated the first peak of $g_{\mathrm{NN}}(r)$ and the main peak of $g_{\mathrm{NH}}(r)$. Among the models that placed the peak correctly, the overestimate of $g_{\mathrm{NN}}(r)$ was largest for the rigid models and smaller for AMOEBA and DPMD, and NQEs lowered it further. NQEs also lowered both peaks of $g_{\mathrm{HH}}(r)$ and moved the inner one slightly outward. The intermolecular RDFs thus separated the models into three groups. The flexible non-polarizable models misplaced the main peaks of $g_{\mathrm{NN}}(r)$ and $g_{\mathrm{NH}}(r)$ and added a short-distance peak to $g_{\mathrm{NH}}(r)$, so their shapes departed most from the experiment. The rigid models placed every main peak correctly but gave the highest $g_{\mathrm{NN}}(r)$ peak of the models that did so. AMOEBA and DPMD followed the experimental curves most closely, with DPMD slightly closer in every peak height (Table~S2), and TRP-DPMD came closest of the models with correctly placed peaks, apart from the inner peak of $g_{\mathrm{HH}}(r)$ that all three share and the experiment lacks.

The models differed most between 2 and 3~\unit{\angstrom} in $g_{\mathrm{NH}}(r)$ (\cref{fig:rdf_inter}c). The rigid models showed a pronounced shoulder there. AMOEBA, DPMD, and TRP-DPMD showed a weaker one, shallowest for TRP-DPMD, and OPLS-AA and OpenFF showed a distinct peak at a shorter distance. This region coincides with the HB distance, so we examined hydrogen bonding next.

\begin{figure}[tbp]
    \centering
    \includegraphics[width=3.37in]{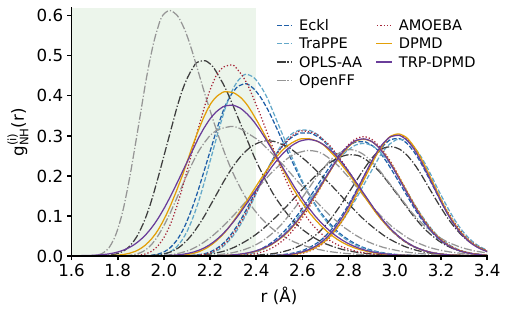}
    \caption{Incremental probability density function, $g^{(i)}_\mathrm{NH}(r)$, of the distance between a nitrogen atom and its $i$-th nearest hydrogen atom in liquid ammonia (see Methods). The distributions correspond to neighbor ranks $i=4,5,6,7$. The shaded area marks $r_{\mathrm{NH}} < 2.4$~\unit{\angstrom}, the hydrogen bond criterion\cite{krishnamoorthy2022hydrogen}.}
    \label{fig:rdfi}
\end{figure}

The differences among the models in $g_{\mathrm{NH}}(r)$ between 2 and 3~\unit{\angstrom} reflect differences in short-range intermolecular coordination. In ammonia, the HB has been defined by the intermolecular \ce{N\bond{-}H} distance alone, $r_{\mathrm{NH}} < 2.4$~\unit{\angstrom}\cite{krishnamoorthy2022hydrogen}, whereas the criteria used for water combine a distance with an angle\cite{luzar1996hydrogen,kumar2007hydrogen}. We adopted this criterion, whose cutoff falls within this region, so that the hydrogen bonding of ammonia can be read directly from the intermolecular \ce{N\bond{-}H} distance distribution. To resolve this distribution by neighbor, we computed the incremental probability density function, $g^{(i)}_{\mathrm{NH}}(r)$\cite{sharma2018born,sharma2018nature,zhuang2022hydration,avula2023understanding,kim2024anomalous}, the distribution of the distance from a nitrogen atom to its $i$-th nearest hydrogen atom. The first three hydrogen atoms belong to the same molecule, so their distributions describe the intramolecular geometry. The fourth is the nearest hydrogen atom of a neighboring molecule, and \cref{fig:rdfi} therefore shows $g^{(i)}_{\mathrm{NH}}(r)$ from $i = 4$ to $7$, where the HB distribution is found. The fourth-neighbor peak characterizes the preferred distance and spatial localization of the nearest intermolecular \ce{N\bond{-}H} contact.

The HB distributions separated the models into the same three groups as the intermolecular RDFs (\cref{fig:rdfi}). In position, AMOEBA, DPMD, and TRP-DPMD coincided for every hydrogen atom from the fourth to the seventh. The flexible non-polarizable models OPLS-AA and OpenFF shifted all four distributions to shorter distances, whereas the rigid models shifted only the fourth, to longer distances. The shorter \ce{N\bond{-}H} separations in OPLS-AA and OpenFF accompanied the inward shift of the first $g_{\mathrm{NN}}(r)$ peak. In height, the models differed most for the fourth hydrogen. The fourth-neighbor peak was highest for OpenFF and lowest for DPMD among the classical nuclei calculations, with TRP-DPMD reducing it further. This ordering indicates progressively less sharply localized nearest-neighbor contacts. The longer contacts of the rigid models appeared as the shoulder of $g_{\mathrm{NH}}(r)$ between 2 and 3~\unit{\angstrom}, and the shorter contacts of OPLS-AA and OpenFF as the distinct peak there.

The models differ more strongly in intermolecular coordination than in the positions of their intramolecular peaks. DPMD captures the principal \ce{N\bond{-}H} and \ce{N\bond{-}N} intermolecular features reasonably well, although discrepancies in peak heights remain. We next assess how these structural differences are reflected in the density and transport properties (\cref{fig:prop}).

\subsubsection{Density and transport properties}
\begin{figure*}[tbp]
    \centering
    \includegraphics[width=6.69in]{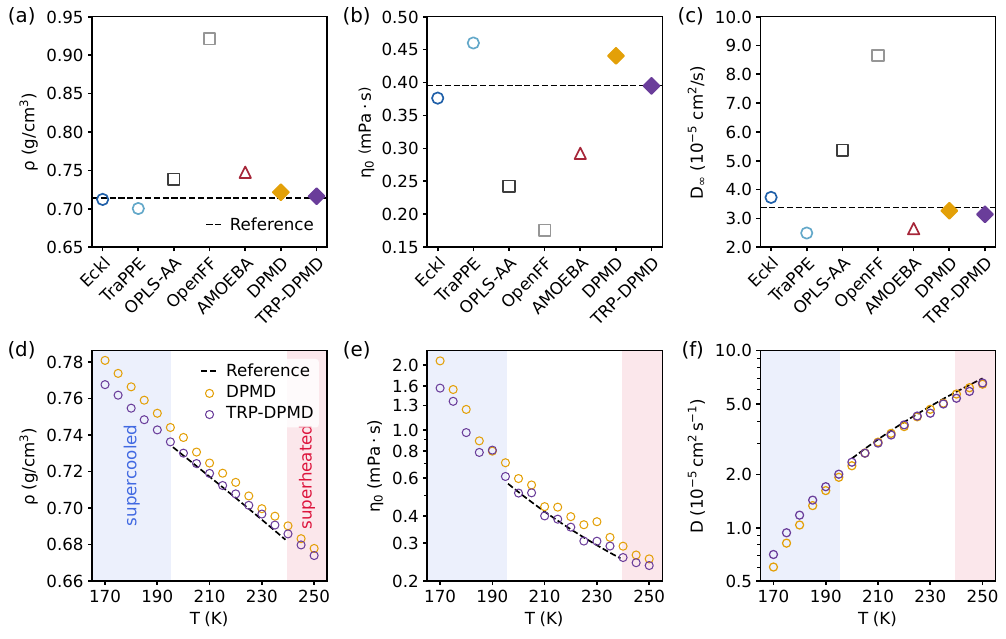}
    \caption{Density, $\rho$, viscosity, $\eta_0$, and diffusion coefficient, $D$, of liquid ammonia. (a--c) Values at 213~\unit{K} for all models. (d--f) Temperature dependence from 170 to 250~\unit{K} for DPMD and TRP-DPMD. The density and transport properties of TRP-DPMD were not computed at 213~\unit{K}, so its values in (a--c) were interpolated between the 210 and 215~\unit{K} points of (d--f). Dashed lines show NIST reference values for $\rho$ and $\eta_0$ and the Arrhenius fit to the NMR diffusion data\cite{oreilly1973self}. Shading marks temperatures below the experimental melting point and above the experimental boiling point at 1~\unit{bar}. In (a--c), circles denote the rigid models Eckl and TraPPE, squares the flexible non-polarizable models OPLS-AA and OpenFF, triangles AMOEBA, and filled diamonds DPMD and TRP-DPMD. Model labels are the same as in \cref{fig:rdf_intra}.}
    \label{fig:prop}
\end{figure*}

We next examine the accuracy of the force fields for the density and transport properties (\cref{fig:prop}a--c). Among the classical nuclei calculations at 213~\unit{K}, DPMD gave the closest agreement with the reference diffusion coefficient\cite{oreilly1973self} and, after Eckl, the second closest agreement with the NIST density and viscosity\cite{nist}, lying slightly above the reference density and about a tenth above the reference viscosity. AMOEBA, by contrast, underestimated both the viscosity and the diffusion coefficient. The flexible non-polarizable models deviated most, with OpenFF overestimating the density by almost a third and the diffusion coefficient by more than a factor of two. Eckl, in turn, matched the density and viscosity most closely, although it overestimated the first peak of $g_{\mathrm{NN}}(r)$ and shifted the nearest intermolecular \ce{N\bond{-}H} contacts to longer distances (\cref{fig:rdf_inter,fig:rdfi}). Its rigid geometry, however, excludes the pyramidal inversion, as shown below.

Over most of the stable liquid interval covered by the reference data, including NQEs improved the density and viscosity while leaving diffusion only weakly affected (\cref{fig:prop}d--f). TRP-DPMD was less dense than DPMD at every temperature studied, and the difference grew as the temperature decreased, so that NQEs removed the excess density of DPMD and TRP-DPMD reproduced the NIST density about as closely as Eckl (\cref{fig:prop}d). Over the liquid range, DPMD was consistently more viscous than the NIST reference, whereas TRP-DPMD agreed with the reference to within a few percent (\cref{fig:prop}e). The diffusion coefficients of TRP-DPMD and DPMD differed by only a few percent in the liquid range, so TRP-DPMD stayed as close to the reference as DPMD there (\cref{fig:prop}f).

Taken together, the structural and bulk-property benchmarks support the MLFF as a consistent description of liquid ammonia across the observables examined here.

\subsection{Hydrogen bonding and local structure}
We next examine how cooling and NQEs modify local HB coordination and packing. Two-state descriptions of water relate its anomalies to the changing populations of locally distinct environments\cite{russo2014understanding,tanaka2020liquid}. This provides a useful structural contrast for ammonia, whose HB coordination and local-environment distributions are examined below.

\begin{figure*}[tbp]
    \centering
    \includegraphics[width=6.69in]{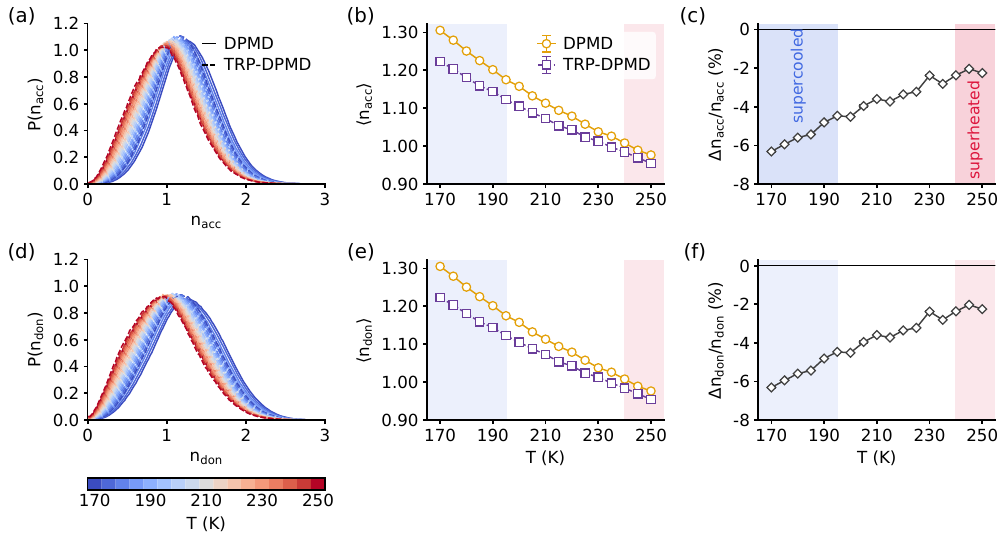}
    \caption{Hydrogen bonding in liquid ammonia from 170 to 250~\unit{K}. (a) Distribution of the number of hydrogen bonds accepted per molecule, $n_{\mathrm{acc}}$, for DPMD (solid) and TRP-DPMD (dashed), with color denoting temperature. (b) Mean $\langle n_{\mathrm{acc}} \rangle$ and (c) relative difference between TRP-DPMD and DPMD, $\Delta n_{\mathrm{acc}}/n_{\mathrm{acc}}$. (d--f) Same for the number of hydrogen bonds donated per molecule, $n_{\mathrm{don}}$. Hydrogen bonds are defined by $r_{\mathrm{NH}} < 2.4$~\unit{\angstrom}\cite{krishnamoorthy2022hydrogen} (see Methods), and the curves in (a) and (d) are smoothed distributions of the integer counts. Blue and red shading mark the supercooled and superheated regions.}
    \label{fig:hist_hb}
\end{figure*}

We first examine the number of HBs per molecule (\cref{fig:hist_hb}). The curves show smoothed distributions of the integer HB counts obtained with the criterion $r_{\mathrm{NH}} < 2.4$~\unit{\angstrom}. The distributions of accepted, $n_{\mathrm{acc}}$, and donated, $n_{\mathrm{don}}$, HBs were unimodal at every temperature and shifted to smaller values with increasing temperature. The mean numbers of accepted and donated HBs were identical, since each HB contributes once to each count, and decreased from 1.30 at 170~\unit{K} to 0.97 at 250~\unit{K} in DPMD. This equality concerns occupied contacts and does not imply balanced numbers of available donor and acceptor sites. TRP-DPMD gave fewer HBs than DPMD at every temperature. The relative difference grew with decreasing temperature, from about 2\% at 250~\unit{K} to 6\% at 170~\unit{K}. TRP-DPMD therefore had lower mean HB coordination under the adopted criterion. This reduction accompanied its lower density and is consistent with a modest loosening of local association.

\begin{figure*}[tbp]
    \centering
    \includegraphics[width=6.69in]{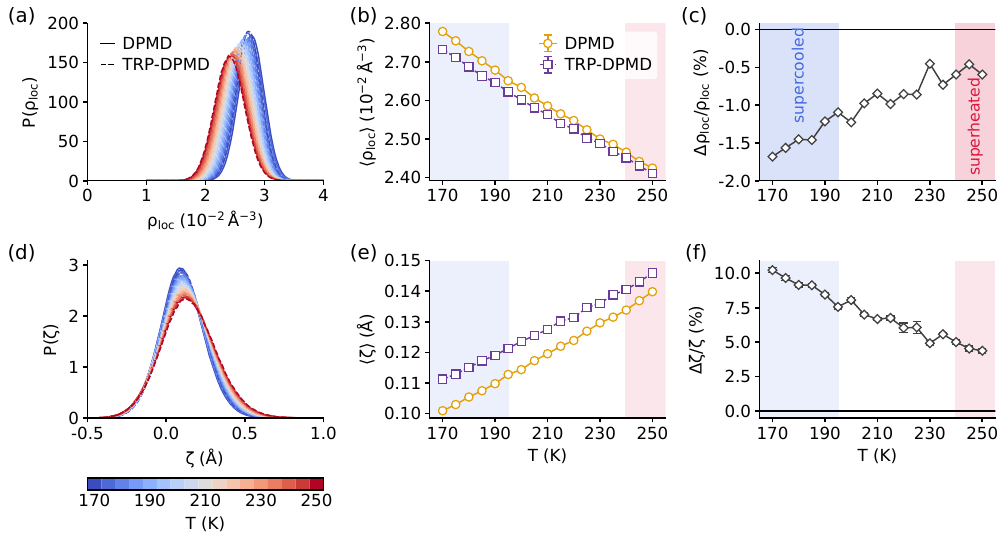}
    \caption{Local density and structural index of liquid ammonia from 170 to 250~\unit{K}. (a) Distribution of the local density, $\rho_{\mathrm{loc}}$, for DPMD (solid) and TRP-DPMD (dashed), with color denoting temperature. (b) Mean $\langle\rho_{\mathrm{loc}}\rangle$ and (c) relative difference between TRP-DPMD and DPMD, $\Delta\rho_{\mathrm{loc}}/\rho_{\mathrm{loc}}$. (d--f) Same for the local structural index, $\zeta$. $\rho_{\mathrm{loc}}$ is the inverse Voronoi volume of the nitrogen atom, and $\zeta$ is the difference between the \ce{N\bond{-}N} distances to the nearest non-hydrogen-bonded and the farthest hydrogen-bonded neighbor (see Methods). Blue and red shading mark the supercooled and superheated regions.}
    \label{fig:hist_prop}
\end{figure*}

We characterized local packing by the inverse Voronoi volume, $\rho_{\mathrm{loc}}$, and the separation of HB and non-HB neighbors by the local structural index, $\zeta$. The latter is the distance to the nearest non-HB neighbor minus that to the farthest HB neighbor\cite{russo2014understanding}. It is evaluated only for molecules with at least one HB neighbor. In two-state descriptions of water, larger $\zeta$ identifies environments with a more clearly separated second shell, whereas values near zero correspond to interpenetrating HB and non-HB coordination shells. The changing populations of these environments provide a structural account of water's anomalies\cite{russo2014understanding,tanaka2020liquid}.

In water, $P(\zeta)$ is bimodal, with one peak near 0~\unit{\angstrom} for a collapsed second shell and another near 1.0~\unit{\angstrom} for a well-ordered second shell\cite{tanaka2019revealing}. The temperature dependence of the two populations reproduces the density and compressibility anomalies\cite{russo2014understanding}, the fragile-to-strong transition of supercooled water\cite{shi2018origin}, and the two-state signature in the scattering function\cite{shi2020direct}. The degree of tetrahedrality controls how strongly these anomalies appear\cite{russo2018waterlike}.

The distributions of the two descriptors were unimodal at every temperature in both DPMD and TRP-DPMD (\cref{fig:hist_prop}a and d). The local density did not separate into the low-density and high-density populations of water\cite{russo2014understanding,tanaka2020liquid}. The distribution $P(\zeta)$ showed a single peak at about 0.1~\unit{\angstrom}, the value of the collapsed second shell in water, and no second population near 1~\unit{\angstrom} appeared even on cooling into the supercooled region. Neither descriptor resolved a separate structural population over the investigated temperature range.

The mean local density decreased monotonically with increasing temperature, whereas the mean $\zeta$ increased (\cref{fig:hist_prop}b and e). NQEs lowered the local density at every temperature, most strongly at the lowest, following the bulk density (\cref{fig:hist_prop}c, \cref{fig:prop}d), as expected for a hydrogen-bonded liquid\cite{ugur2025nuclear}. They raised $\zeta$ by several percent (\cref{fig:hist_prop}f), but the increase left it far below the value of the ordered second shell of water. NQEs reduced the mean HB coordination and local density while producing only a small absolute increase in $\zeta$. The distributions remained unimodal, indicating that these shifts occur without a resolved separation into distinct local environments.

\subsection{Nuclear quantum effects on diffusion}
\begin{figure}[tbp]
    \centering
    \includegraphics[width=3.37in]{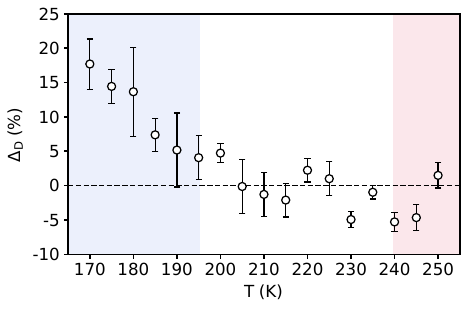}
    \caption{Relative difference between the diffusion coefficients of TRP-DPMD and DPMD, $\Delta_D = (D_{\mathrm{TRP\text{-}DPMD}} - D_{\mathrm{DPMD}})/D_{\mathrm{DPMD}}$, as a function of temperature. Both coefficients are averages over independent simulations, seven for DPMD and two for TRP-DPMD, and were corrected for finite-size effects\cite{yeh2004system} (see Methods). Error bars denote the uncertainty in $\Delta_D$, propagated from the standard errors of the independently estimated diffusion coefficients. The dashed line marks $\Delta_D = 0$. Blue and red shading mark the supercooled and superheated regions as in \cref{fig:prop}.}
    \label{fig:anomalous_D}
\end{figure}

NQEs affected the diffusion coefficient differently from the density, the viscosity, the number of HBs, and the local density. For those quantities, the effect was a decrease at every temperature studied, and only its size varied with temperature, growing toward low temperatures (\cref{fig:prop}d and e, \cref{fig:hist_hb}c and f, and \cref{fig:hist_prop}c). For the diffusion coefficient, the sign of the effect itself depended on temperature (\cref{fig:anomalous_D}). We quantified the effect by the relative difference, $\Delta_D = (D_{\mathrm{TRP\text{-}DPMD}} - D_{\mathrm{DPMD}})/D_{\mathrm{DPMD}}$. This comparison follows the 1~\unit{bar} isobar and therefore includes the accompanying change in equilibrium density. The reported diffusion coefficients also include the viscosity-dependent finite-size correction in \cref{eq:yhrelation}. Between 210 and 250~\unit{K}, $\Delta_D$ stayed between $-5$ and $+2$\%, with slightly negative central estimates at most temperatures. It became positive on cooling below approximately 210~\unit{K} and reached 18\% at 170~\unit{K}. The crossover thus concerns the quantum contribution to diffusion rather than an increase of the diffusion coefficient upon cooling. We next examine pyramidal inversion and its coupling to local relaxation as a microscopic basis for this temperature dependence.

\subsection{Pyramidal inversion in the liquid}
\begin{figure}[tbp]
    \centering
    \includegraphics[width=3.37in]{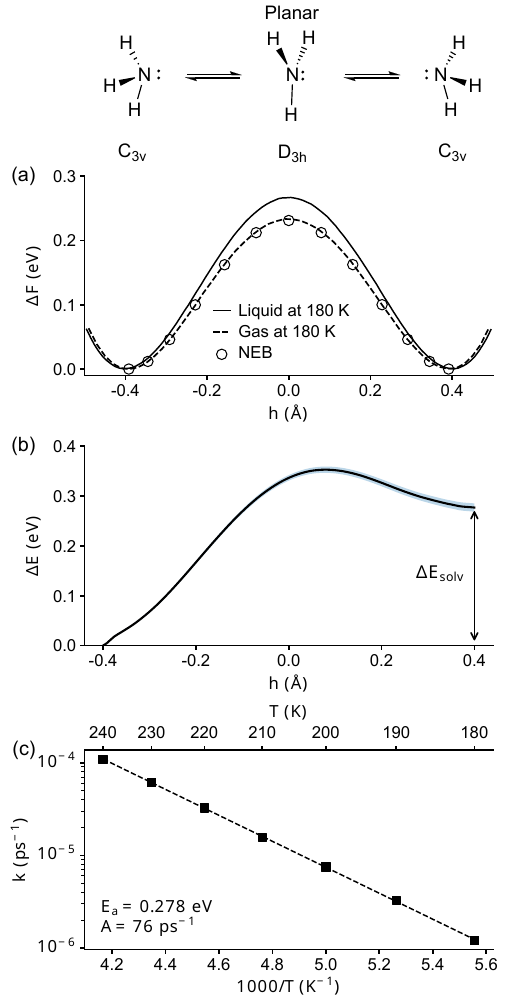}
    \caption{Pyramidal inversion of ammonia. (a) Free energy profiles of the inversion along the inversion coordinate, $h$, in liquid ammonia (solid line) and in the gas (dashed line) at 180~\unit{K}, obtained with DPMD coupled with enhanced sampling, together with the minimum energy path of an isolated molecule from a nudged elastic band (NEB) calculation (circles). The schematic above the panel shows the two pyramidal configurations and the planar transition state. (b) Energy along $h$ from an NEB calculation in the liquid in which all molecules except the inverting molecule were held fixed. $\Delta E_{\mathrm{solv}}$ marks the energy difference between the inverted and the initial configuration. (c) Arrhenius plot of the inversion rate constant, $k$, estimated by transition state theory from the liquid free energy profiles at 180--240~\unit{K}. The dashed line is the fit to \cref{eq:arrhenius} with $E_a = 0.278$~\unit{eV} and $A = 76$~\unit{ps^{-1}}.}
    \label{fig:inversion}
\end{figure}

One of the distinctive molecular motions of ammonia is the pyramidal inversion, in which the nitrogen atom passes through the plane of the three hydrogen atoms to the opposite side (top of \cref{fig:inversion})\cite{nguyen2021large}. For an isolated molecule, the two pyramidal configurations are equivalent minima of a symmetric double-well potential\cite{spirko1983vibrational,rajamaki2003vibrational}. We describe the inversion by the inversion coordinate, $h$, the signed distance of the nitrogen atom from the plane of its three hydrogen atoms, so that the two pyramidal minima lie at $h \approx \pm 0.4$~\unit{\angstrom} and the planar transition state at $h = 0$. In the gas phase, the lowest vibrational levels lie below this barrier and the molecule inverts by tunneling\cite{flambaum2007limit,nguyen2021large}. The inversion is sensitive to the environment. The gas-phase inversion resonance broadens and shifts to lower frequency with increasing pressure, becoming quenched near 1.7~\unit{bar}\cite{herbauts2007quantum}. This pressure dependence demonstrates the sensitivity of the coherent inversion response to intermolecular interactions.

Solvation raises the classical free energy barrier to inversion in the MLFF. We computed the free energy profiles of the inversion in liquid and gaseous ammonia at 180~\unit{K} with DPMD coupled with enhanced sampling (see Methods) and compared them with the minimum energy path from an NEB calculation of an isolated molecule (\cref{fig:inversion}a). The free energy barrier in the gas coincided with the NEB energy barrier, whereas the barrier in the liquid at 180~\unit{K} was about 0.03~\unit{eV} higher than both, consistent with suppressed thermally activated inversion in the condensed environment.

The energetic role of the surrounding liquid is illustrated by an inversion path calculated with the neighboring molecules held fixed. To separate the inversion from the relaxation of the solvation shell, we performed an NEB calculation of the inversion in the liquid in which all molecules except the inverting one were held fixed at their positions in the initial configuration (\cref{fig:inversion}b). The path started from the pyramidal minimum at $h = -0.4$~\unit{\angstrom} and ended at the inverted configuration at $h = +0.4$~\unit{\angstrom}. With the surroundings frozen, the energy barrier was 0.352~\unit{eV}, higher than the free energy barrier obtained with mobile surroundings (\cref{fig:inversion}a), and the inverted configuration lay 0.277~\unit{eV} above the initial one. The large energy asymmetry shows that the initial solvation environment poorly accommodates the inverted geometry. Together with the relaxed free energy profile, this result establishes an energetic coupling between inversion and solvent reorganization.

The transition state theory rate estimates obtained from the free energy profiles followed an Arrhenius dependence from 180 to 240~\unit{K} (\cref{fig:inversion}c). We computed the free energy profile of the inversion for 512 \ce{NH3} molecules in the NVT ensemble from 180 to 240~\unit{K} at the corresponding densities of \cref{fig:prop}d, using DPMD coupled with enhanced sampling. From each profile we obtained the rate constant, $k$, by transition state theory. Fitting these estimates to \cref{eq:arrhenius} gave an apparent activation energy, $E_a$, of 0.278~\unit{eV} and a prefactor, $A$, of 76~\unit{ps^{-1}}. The Arrhenius equation reads
\begin{equation}
  k = A\exp\!\left(-\frac{E_a}{k_{\mathrm{B}}T}\right).
  \label{eq:arrhenius}
\end{equation}

\begin{figure*}[tbp]
    \centering
    \includegraphics[width=6.69in]{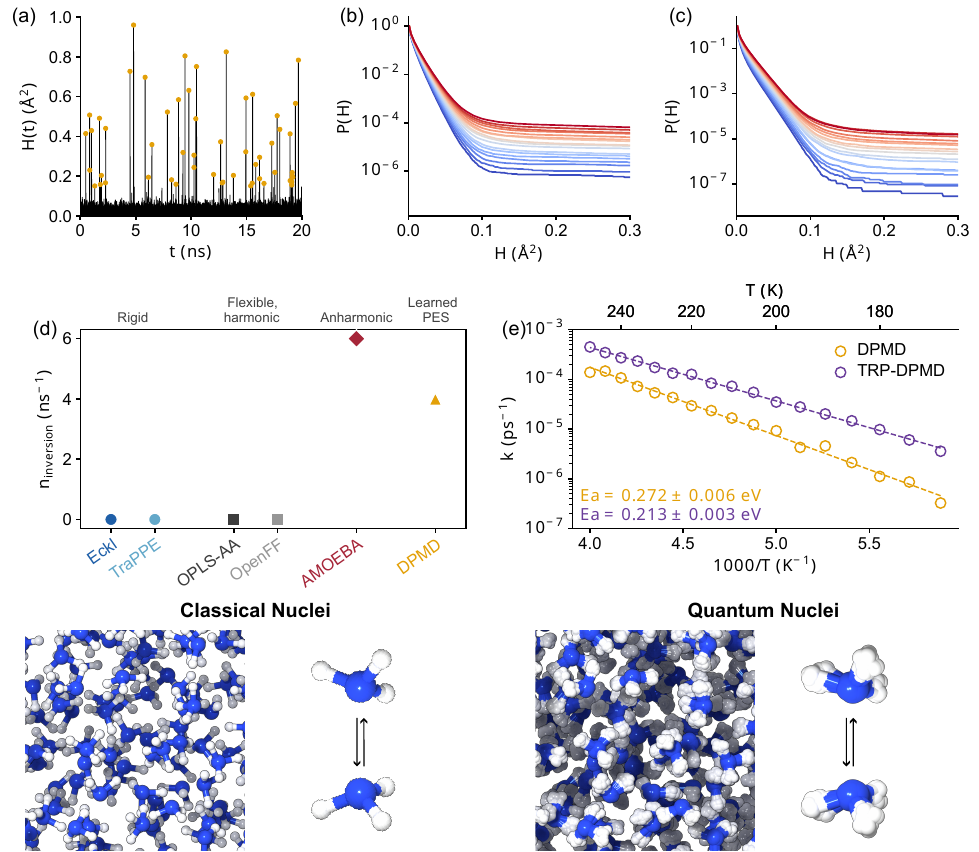}
    \caption{Detection and counting of inversion events in MD simulations. (a) Hop function, $H$, of the inversion coordinate, $h$, at 215~\unit{K}. Spikes correspond to inversion events. (b, c) Exceedance probability, $P(H)$, that the hop function at a time step exceeds the value $H$, for (b) TRP-DPMD and (c) DPMD from 170 to 250~\unit{K}. (d) Number of inversion events per nanosecond in each force field at 213~\unit{K}, grouped as rigid, flexible with harmonic terms, anharmonic, and learned potential energy surface. (e) Arrhenius plot of the inversion rate constant, $k$, from event counting for DPMD (orange) and TRP-DPMD (purple). Dashed lines are Arrhenius fits with the indicated activation energies, $E_a$. The bottom row shows snapshots of the liquid and of the two pyramidal configurations of an inverting molecule with classical nuclei (left) and with quantum nuclei represented by ring polymers (right).}
    \label{fig:inversion_MD}
\end{figure*}

The inversion also occurs spontaneously within the nanosecond time scale of the unbiased MD simulations, so it can be observed directly without enhanced sampling. To count inversion events in each force field, we used the hop function, $H(t)$, which was introduced to detect cage jumps of grains in dense granular media\cite{candelier2009building} and has since been used to identify discrete ion hops in organic ionic plastic crystals\cite{park2026beyond}. Evaluated on the inversion coordinate, $h$, it compares the values of $h$ in two adjacent time windows, one before and one after $t$, and becomes large when $h$ changes abruptly between them. \Cref{fig:inversion_MD}a shows $H(t)$ in a DPMD simulation at 215~\unit{K} in the NVT ensemble. The sharp spikes correspond to inversion events, during which $h$ changes sign within a short time. With $\langle\cdot\rangle_{-}$ and $\langle\cdot\rangle_{+}$ denoting averages over the windows before and after $t$, the hop function is defined as \cref{eq:hop}.
\begin{equation}
    H(t) = \sqrt{
    \left\langle \left( h(t) - \left\langle h(t) \right\rangle_{+} \right)^2 \right\rangle_{-}
    \left\langle \left( h(t) - \left\langle h(t) \right\rangle_{-} \right)^2 \right\rangle_{+}
    }
  \label{eq:hop}
\end{equation}

To set a threshold for counting inversion events, we examined the exceedance probability, $P(H)$, that the hop function at a time step exceeds a given value, for TRP-DPMD (\cref{fig:inversion_MD}b) and DPMD (\cref{fig:inversion_MD}c) at each temperature. The probability $P(H)$ decreased steeply up to about 0.1~\unit{\angstrom^2} and then changed slope abruptly. Values below 0.1~\unit{\angstrom^2} therefore arise from ordinary vibrations of $h$ about a pyramidal minimum, whereas larger values arise from inversions. We set the threshold at 0.15~\unit{\angstrom^2} to stay clear of the crossover.

In the classical benchmark simulations, spontaneous pyramidal inversion was observed only with AMOEBA and DPMD. We counted inversion events in the simulations of \cref{fig:rdf_intra} at 213~\unit{K} (\cref{fig:inversion_MD}d). Eckl, TraPPE, OPLS-AA, and OpenFF showed no inversion, whereas AMOEBA and DPMD yielded approximately 6 and 4 inversion events per nanosecond, respectively, in the 512 \ce{NH3} molecule simulation. The per-molecule rate constant used below is $k = N_{\mathrm{inv}}/(N_{\mathrm{mol}}\,t_{\mathrm{obs}})$, with $N_{\mathrm{inv}}$ the number of events, $N_{\mathrm{mol}}$ the number of molecules, and $t_{\mathrm{obs}}$ the observation time. The rigid-body potentials, Eckl and TraPPE, constrain the motion of the hydrogen atoms in ammonia to fixed bond lengths and angles. The flexible non-polarizable models OPLS-AA and OpenFF describe the angular motion with a harmonic potential centered at 106.4\unit{\degree} and 110.4\unit{\degree}, respectively.\cite{rizzo1999opls,boothroyd2023development} At the planar transition state of the inversion, the \ce{H\bond{-}N\bond{-}H} angle of ammonia is 120\unit{\degree}. Rigid geometries exclude inversion by construction. In OPLS-AA and OpenFF, approaching the planar geometry incurs an angular energy penalty, and no inversion was sampled over the present trajectories. AMOEBA uses an anharmonic valence description, while the MLFF represents the energy landscape without imposing a prescribed harmonic angular form.

We obtained the apparent activation energies by fitting the per-molecule inversion rate constants from the MD simulations to \cref{eq:arrhenius} (\cref{fig:inversion_MD}e). For DPMD, the activation energy from event counting, $0.272 \pm 0.006$~\unit{eV}, agreed with the value of 0.278~\unit{eV} obtained from the free energy profiles of the enhanced-sampling simulations. TRP-DPMD reduced the apparent activation energy to $0.213 \pm 0.003$~\unit{eV} (\cref{fig:inversion_MD}e). The smaller magnitude of the Arrhenius slope shows that inversion is less strongly suppressed by cooling with quantum nuclei. This change occurs on the same underlying MLFF potential energy surface and is consistent with quantum fluctuations modifying the coupled inversion and solvation dynamics. Both absolute rates decreased on cooling, but the relative enhancement in TRP-DPMD became larger.

In summary, DPMD provided a consistent description of the liquid structure, density, viscosity, and diffusion coefficient of ammonia and was one of only two models to sample spontaneous pyramidal inversion in the classical benchmark simulations. DPMD is therefore suitable for studying the inversion in liquid ammonia. In the following sections, we use DPMD and TRP-DPMD for a detailed analysis of the inversion and its relation to the temperature dependence of the quantum correction to diffusion.

\subsection{Structural perturbation by the inversion}
To investigate how the local structural properties change during an inversion event, we calculated the local properties around the inversion time. We analyzed inversion events sampled in six independent 10.0~\unit{ns} DPMD simulations in the NVT ensemble at 240~\unit{K}. We first examine the impact of the inversion on the inverting ammonia molecule itself and then on the neighboring molecules.

\begin{figure}[tbp]
    \centering
    \includegraphics[width=3.37in]{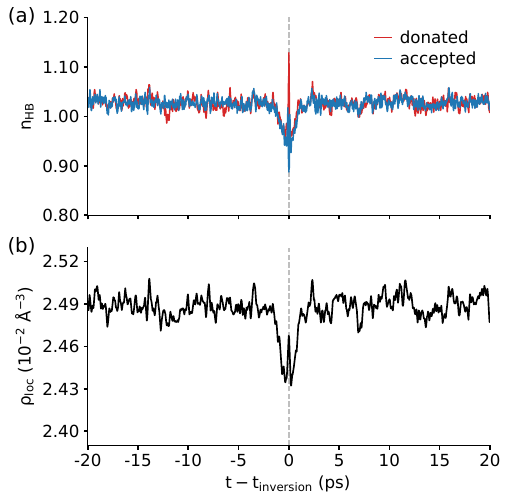}
    \caption{Structural changes of the inverting molecule around inversion events in DPMD simulations at 240~\unit{K}. (a) Number of hydrogen bonds, $n_{\mathrm{HB}}$, donated by the inverting molecule and accepted by it, and (b) local density, $\rho_{\mathrm{loc}}$, of the inverting molecule, as a function of time relative to the inversion event, $t - t_{\mathrm{inversion}}$. The dashed line marks $t = t_{\mathrm{inversion}}$. Values are averages over all inversion events (see Methods).}
    \label{fig:7}
\end{figure}

Inversion events were accompanied by a transient depletion of both accepted and donated HBs (\cref{fig:7}a). The depletion developed approximately 1~\unit{ps} before the event and relaxed over a comparable time afterward. The accepted count reached a minimum near inversion, while the donated count exhibited a narrow positive spike superimposed on the broader depletion. The spike is consistent with transient close contacts generated as the molecule approaches the planar geometry, in which the \ce{H\bond{-}N\bond{-}H} angle opens to 120\unit{\degree}, sweeping the hydrogen atoms outward, close to the nitrogen atoms of neighboring molecules. Under the distance based criterion, these contacts increase the apparent donated coordination without necessarily indicating strengthened HB association.

The inversion also perturbed the local density of the inverting molecule, which dipped over the same window and recovered afterward (\cref{fig:7}b). The coincident reduction in local density indicates a transient loosening of the environment around the inverting molecule. Together with HB depletion, it provides a structural signature of the solvation reorganization suggested by the frozen environment calculation in \cref{fig:inversion}b. The temperature resolved analysis in Fig.~S1 shows reduced HB coordination and local density at inversion throughout the investigated range of 170--250~\unit{K} for TRP-DPMD. The same tendency is resolved for DPMD except at the three lowest temperatures, where too few inversion events were sampled to resolve the changes. These observations support the interpretation of inversion as a local structure breaker coupled to solvation-shell relaxation.

\begin{figure}[tbp]
    \centering
    \includegraphics[width=3.37in]{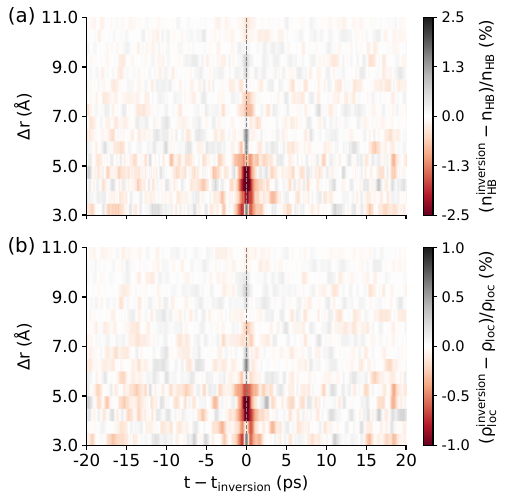}
    \caption{Propagation of the structural perturbation from an inverting molecule to its surroundings at 240~\unit{K}. Relative change of (a) the number of hydrogen bonds, $n_{\mathrm{HB}}$, and (b) the local density, $\rho_{\mathrm{loc}}$, of the molecules at a distance $\Delta r$ from the inverting molecule, as a function of time relative to the inversion event, $t - t_{\mathrm{inversion}}$. The dashed line marks $t = t_{\mathrm{inversion}}$.}
    \label{fig:impact_others}
\end{figure}

To examine whether the inversion perturbation extends to the neighboring molecules, we resolved the relative change of the number of HBs and of the local density by the distance from the inverting molecule and by the time relative to the inversion (\cref{fig:impact_others}). On average over events, the molecules within about 5.0~\unit{\angstrom} of the inverting molecule also lost HBs and local density at the inversion time, by about 2\% and 1\%, respectively. This distance, 5.0~\unit{\angstrom}, corresponds to the first minimum of $g_{\mathrm{NN}}(r)$, that is, the boundary of the first solvation shell (\cref{fig:rdf_inter}b). No corresponding signal was resolved beyond that shell in either descriptor, and the local response decayed within a few picoseconds. This recovery is slower than the reported lifetime of an individual HB, about 0.1~\unit{ps}\cite{krishnamoorthy2022hydrogen}, which suggests that the response involves reorganization of the local coordination environment rather than the persistence of a single bond.

The short range response is consistent with the sparse and transient HB connectivity of ammonia. Each nitrogen atom accepts about one HB (\cref{fig:hist_hb}), and these bonds are short lived. Unlike a persistent network, this connectivity offers limited scope for transmitting a coordinated HB rearrangement over larger distances. In water, by contrast, each oxygen atom accepts two\cite{krishnamoorthy2022hydrogen}, forming a densely connected network in which HB rearrangements are collective\cite{ohmine1993fluctuation,fecko2005local,nicodemus2011collective,schulz2018collective}.

\subsection{Inversion and cage escape}
The preceding results connect inversion with localized structural disruption and show that quantum nuclei enhance its rate. The enhancement of diffusion below approximately 210~\unit{K} motivates examining how this internal motion relates to translational relaxation. We therefore compare inversion-conditioned displacements and the probability of displacements beyond the cage length.

\begin{figure*}[tbp]
    \centering
    \includegraphics[width=6.69in]{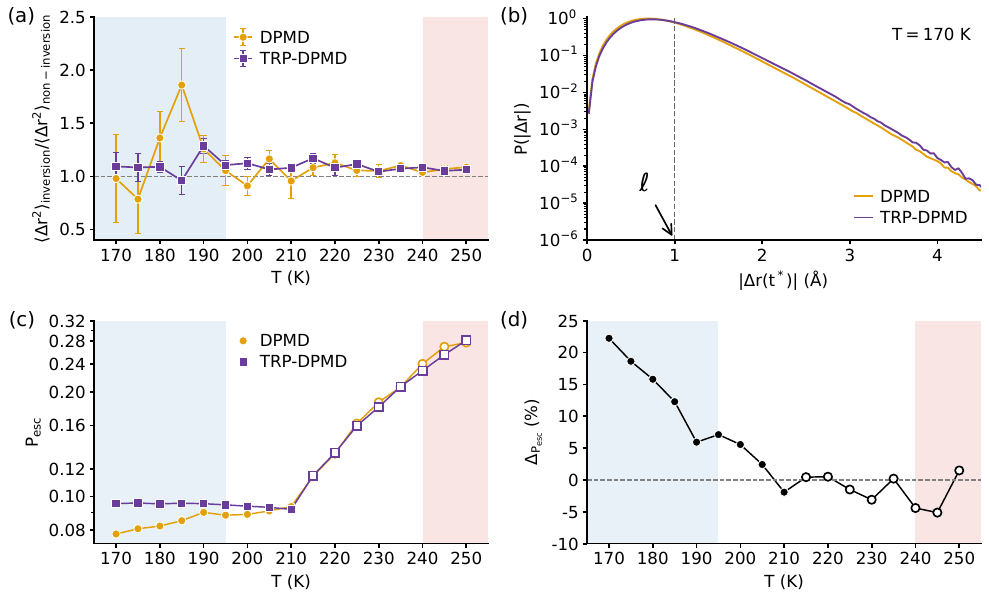}
    \caption{Displacement of inverting molecules and cage escape. (a) Ratio of the mean square displacement, $\langle\Delta r^{2}\rangle$, of molecules that inverted within the time window to that of molecules that did not. (b) Distribution of the displacement over $t^{*}$, the time at which the logarithmic slope of $\langle\Delta r^{2}\rangle$ is minimal, at 170~\unit{K}. The dashed line marks the cage length, $\ell$. (c) Cage-escape probability, $P_{\mathrm{esc}}$, that the displacement over $t^{*}$ exceeds $1.5\,\ell$. Filled symbols denote temperatures at which a cage was resolved. Open symbols use the $t^{*}$ and $\ell$ of 210~\unit{K}, the highest such temperature. (d) Relative difference of this probability between TRP-DPMD and DPMD. Blue and red shading mark the supercooled and superheated regions.}
    \label{fig:diffusion}
\end{figure*}

Molecules that inverted generally showed larger nitrogen displacements than molecules that did not. We computed the mean square displacement, $\langle\Delta r^{2}\rangle$, of the nitrogen atoms separately for the two groups over the same time window and compared them (\cref{fig:diffusion}a, see Methods). The ratio was close to 1.1 at most temperatures in both DPMD and TRP-DPMD, with larger scatter in the low temperature DPMD results. This $\langle\Delta r^{2}\rangle$ contains both the translation of the molecule and the bounded intramolecular motion of the nitrogen atom through the plane of the hydrogen atoms during the inversion (\cref{fig:inversion_MD}a), so the ratio does not by itself measure translational diffusion. The accompanying loosening of the solvation environment (\cref{fig:7,fig:impact_others}) provides an additional route to enhanced displacement through cage relaxation, which we examine next at the scale of the cage.

Below about 210~\unit{K}, the cage-escape probability was higher in TRP-DPMD than in DPMD when evaluated using the same observation time and displacement threshold. We defined the observation time, $t^{*}$, from the minimum logarithmic slope of $\langle\Delta r^{2}\rangle$ and the cage length, $\ell = \sqrt{\langle\Delta r^{2}(t^{*})\rangle}$ (see Methods). The cage-escape probability, $P_{\mathrm{esc}} = P(|\Delta r(t^{*})| > 1.5\,\ell)$, is evaluated over the full molecular population. At the lowest temperature, this probability was higher in TRP-DPMD than in DPMD (\cref{fig:diffusion}b). Above approximately 210~\unit{K}, no cage was resolved by the present criterion. With the observation window fixed at its 210~\unit{K} value, $P_{\mathrm{esc}}$ increased rapidly with temperature and was similar in the two calculations (\cref{fig:diffusion}c). Below this crossover, $P_{\mathrm{esc}}$ remained nearly constant in TRP-DPMD over the corresponding temperature-dependent observation windows, whereas it decreased in DPMD. The relative difference reached approximately 20\% at 170~\unit{K} (\cref{fig:diffusion}d). This difference followed the same temperature trend as the enhancement of diffusion by NQEs (\cref{fig:anomalous_D}).

We attribute the temperature trend of the quantum correction to diffusion to the interplay of the cage and the inversion. In the warmer liquid, the quantum correction to diffusion remains small despite the higher inversion rate in TRP-DPMD. As translational dynamics becomes suppressed by caging, a molecule must break out of its cage to move far, and the local coordination loss accompanying inversion (\cref{fig:7,fig:impact_others}) can contribute more strongly to relaxation. With decreasing temperature, the cage deepens and escape becomes rarer in DPMD. Both inversion rates decrease on cooling, but the rate in TRP-DPMD becomes increasingly enhanced relative to DPMD. This contrast supports a cage-dependent transport contribution from quantum-enhanced inversion, rather than the onset of inversion itself at low temperature.

\begin{figure}[tbp]
    \centering
    \includegraphics[width=3.37in]{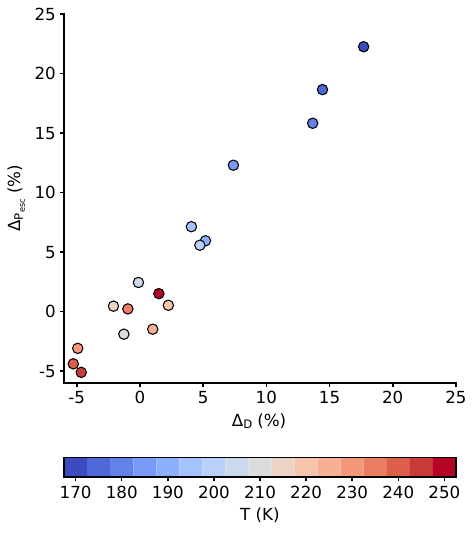}
    \caption{Relation between the enhancement of cage escape and the enhancement of diffusion by NQEs. The vertical axis is the relative difference of the cage-escape probability, $P_{\mathrm{esc}}$, between TRP-DPMD and DPMD, $\Delta_{P_{\mathrm{esc}}} = (P_{\mathrm{esc, TRP\text{-}DPMD}} - P_{\mathrm{esc, DPMD}})/P_{\mathrm{esc, DPMD}}$ (\cref{fig:diffusion}d). The horizontal axis is the relative difference of the diffusion coefficient, $\Delta_D = (D_{\mathrm{TRP\text{-}DPMD}} - D_{\mathrm{DPMD}})/D_{\mathrm{DPMD}}$ (\cref{fig:anomalous_D}). Color denotes temperature.}
    \label{fig:end}
\end{figure}

The relative changes in the cage-escape probability and in the diffusion coefficient by NQEs followed an approximately linear trend across the investigated temperatures (\cref{fig:end}). This correspondence connects the quantum enhancement of long-time diffusion in \cref{fig:anomalous_D} with an increased probability of cage-scale displacements. Neither axis measures the inversion itself, so the correlation alone does not establish the mechanism. Two independent observations, however, tie it to the inversion. Event-conditioned averages show transient disruption of the first solvation shell (\cref{fig:7,fig:impact_others}). The inversion rate is higher with quantum nuclei, and its relative enhancement grows on cooling (\cref{fig:inversion_MD}e). Together, these results identify quantum-enhanced pyramidal inversion as a microscopic contributor to translational relaxation in the caged regime below about 210~\unit{K}.

\section{Conclusions}
We have connected pyramidal inversion to local structural relaxation and quantum-enhanced diffusion in liquid ammonia. NQEs lower the density and viscosity throughout the investigated range, while their contribution to diffusion becomes positive below approximately 210~\unit{K} and reaches 18\% at 170~\unit{K}. The accompanying enhancement of inversion and cage escape supports an inversion-assisted cage-relaxation mechanism as translational dynamics becomes suppressed on cooling.

The previously developed MLFF provides a consistent description of liquid structure and bulk properties while retaining spontaneous pyramidal inversion. Including NQEs improves the density and viscosity over the reference liquid interval. Among the five conventional force fields tested, only AMOEBA sampled inversion during the present trajectories. This comparison highlights the distinction between reproducing bulk properties and retaining the internal molecular pathway relevant to their microscopic interpretation.

The free energy and frozen-environment calculations connect inversion energetics to the arrangement of neighboring molecules. Event conditioned averages show transient HB depletion and reduced local density around inversion events, with no corresponding signal resolved beyond the first solvation shell. NQEs lower the apparent inversion activation energy, while their relative enhancements of cage-escape probability and diffusion follow similar temperature dependences. Together, these observations support an inversion-assisted contribution to relaxation. Because the classical and quantum simulations are compared at their respective equilibrium densities, the comparison does not isolate this contribution from the accompanying density change.

These localized rearrangements coexist with unimodal distributions of the examined structural descriptors. The results thus connect quantum-sensitive intramolecular motion to a local relaxation pathway without requiring a resolved separation into distinct structural populations.

Systematic comparisons across chemical space have related NQEs on liquid thermophysical properties to molecular composition and intermolecular interactions\cite{ugur2025nuclear}. The present results complement this perspective by highlighting a distinction between the quantum sensitivity of an internal motion and its significance for translational relaxation. Inversion is quantum-enhanced even in the warmer liquid, where quantum and classical diffusion coefficients remain similar. On cooling, both absolute inversion rates decrease, but their quantum-classical contrast grows alongside the relative enhancement of cage escape and diffusion. These trends support a picture in which the local rearrangements coupled to inversion become more consequential as neighboring molecules constrain translational motion. The central implication is therefore not simply that quantum nuclei accelerate an internal motion, but that its coupling to local cage relaxation can make that motion relevant to long-time transport.

\section*{Supplementary Material}
See the supplementary material for the peak positions and heights of the radial distribution functions (Tables~S1 and S2) and the temperature dependence of the structural change at the inversion (Fig.~S1).

\begin{acknowledgments}
This work was supported by the National Research Foundation of Korea (NRF) grant funded by the Ministry of Science and ICT (MSIT) of the Korean government (RS-2024-00343871) and by the Technology Innovation Program (RS-2025-19532970) funded by the Ministry of Trade, Industry and Resources (MOTIR, Korea) (W.B.L.); by the NRF grant funded by the Ministry of Education of the Korean government (RS-2021-NR060141) (J.W.Y.); and by the Center for Advanced Computation at the Korea Institute for Advanced Study (KIAS).
\end{acknowledgments}

\section*{Author Declarations}
\subsection*{Conflict of Interest}
The authors have no conflicts to disclose.

\subsection*{Author Contributions}
\textbf{Minwoo Kim:} Conceptualization (equal); Methodology (lead); Software (lead); Investigation (lead); Formal analysis (lead); Visualization (lead); Writing -- original draft (lead); Writing -- review \& editing (equal). \textbf{Hido Woo:} Investigation (supporting); Formal analysis (supporting); Writing -- review \& editing (supporting). \textbf{Taeyong Park:} Investigation (supporting); Formal analysis (supporting); Writing -- review \& editing (supporting). \textbf{Ji Woong Yu:} Conceptualization (equal); Supervision (equal); Funding acquisition (equal); Writing -- review \& editing (equal). \textbf{Won Bo Lee:} Conceptualization (equal); Supervision (equal); Project administration (lead); Funding acquisition (equal); Writing -- review \& editing (equal).

\section*{Data Availability Statement}
The data that support the findings of this study are available from the corresponding authors upon reasonable request.

\bibliography{references}
\end{document}